\documentclass[12pt]{article}
\usepackage{times}
\usepackage{geometry}
\usepackage[utf8]{inputenc}
\usepackage{enumitem,amssymb}
\usepackage{ragged2e}
\usepackage{setspace}
\usepackage{graphicx}
\usepackage{floatrow}
\newlist{thematic}{itemize}{8}
\setlist[thematic]{label=$\square$}
\usepackage{pifont}
\newcommand{\cmark}{\ding{51}}%
\newcommand{\done}{\rlap{$\square$}{\raisebox{2pt}{\large\hspace{1pt}\cmark}}%
\hspace{-2.5pt}}

\begin{document}
\pagenumbering{gobble}
\RaggedRight
\noindent {\fontsize{16}{20} \selectfont Astro2020 Science White Paper}
\begin{center}
{\fontsize{24}{32}\selectfont Probing Magnetic Reconnection
in Solar Flares}
\vspace{0.5cm}

\textit{\fontsize{16}{20}\selectfont New Perspectives from Radio Dynamic Imaging Spectroscopy}
\end{center}

\vspace{0.3cm}

\normalsize

\noindent \textbf{Thematic Areas:} \hspace*{60pt} $\square$ Planetary Systems \hspace*{10pt} $\square$ Star and Planet Formation \hspace*{20pt}\linebreak
$\square$ Formation and Evolution of Compact Objects \hspace*{31pt} $\square$ Cosmology and Fundamental Physics \linebreak
  \done Stars and Stellar Evolution \hspace*{1pt} $\square$ Resolved Stellar Populations and their Environments \hspace*{40pt} \linebreak
  $\square$    Galaxy Evolution   \hspace*{45pt} $\square$             Multi-Messenger Astronomy and Astrophysics \hspace*{65pt} \linebreak

\justifying
  
\noindent \textbf{Principal Author:} \\
Bin Chen, \textit{New Jersey Institute of Technology} \\
Email: bin.chen@njit.edu \\
Phone: (973) 596-3565 \\
\textbf{Co-authors:} \\
Tim Bastian, \textit{National Radio Astronomy Observatory} \\
Joel Dahlin, \textit{NASA Goddard Space Flight Center} \\
James F. Drake, \textit{University of Maryland} \\
Gregory D. Fleishman, \textit{New Jersey Institute of Technology} \\
Dale E. Gary, \textit{New Jersey Institute of Technology} \\
Lindsay Glesener, \textit{University of Minnesota} \\
Fan Guo, \textit{Los Alamos National Laboratory} \\
Hantao Ji, \textit{Princeton University} \\
Pascal Saint-Hilaire, \textit{University of California, Berkeley} \\
Chengcai Shen, \textit{Center for Astrophysics} \textbar\ \textit{Harvard \& Smithsonian}  \\
Stephen M. White, \textit{Air Force Research Laboratory} \\

\noindent \textbf{Executive Summary:}

Magnetic reconnection is a fundamental physical process in many laboratory, space, and astrophysical plasma contexts. Solar flares serve as an outstanding laboratory to study the magnetic reconnection and the associated energy release and conversion processes under plasma conditions difficult to reproduce in the laboratory, and with considerable spatiotemporal details not possible elsewhere in astrophysics. Here we emphasize the unique power of remote-sensing observations of solar flares at radio wavelengths. In particular, we discuss the transformative technique of \textit{broadband radio dynamic imaging spectroscopy} in making significant contributions to addressing several outstanding challenges in magnetic reconnection, including the capability of pinpointing magnetic reconnection sites, measuring the time-evolving reconnecting magnetic fields, and deriving the spatially and temporally resolved distribution function of flare-accelerated electrons. 

\pagebreak
\pagenumbering{arabic}
\noindent \textbf{\large Introduction}

Magnetic reconnection, a fundamental physical process in which magnetic field lines undergo a sudden topological reconfiguration to release magnetic energy, is thought to play a key role in powering explosive flare activities in many astrophysical and space plasma systems. Outstanding examples include terrestrial substorms in the Earth's magnetosphere \cite{2008Sci...321..931A}, solar and stellar flares \cite{2017LRSP...14....2B,1991ARA&A..29..275H}, blazar jets \cite{2013MNRAS.431..355G}, gamma ray bursts \cite{1994MNRAS.270..480T}, pulsar wind nebulae \cite{2017ASSL..446..247S}, and accretion disks around protostars \cite{2000ApJ...532.1097M} and black holes \cite{1998MNRAS.299L..15D}. Magnetic reconnection is also believed as an important mechanism for accelerating charged particles to high energies \cite{1997JGR...10214631M,2011SSRv..159..357Z}, and might be responsible for heating the solar and stellar coronae to multi-million degrees \cite{1988ApJ...330..474P}. By virtue of their proximity, solar flares---the largest explosions in the solar system---serve as an excellent laboratory to study the magnetic-reconnection-driven magnetic energy release and the subsequent energy conversion processes in spatiotemporal detail not possible elsewhere in astrophysics.

\begin{figure}[!ht]
\floatbox[{\capbeside\thisfloatsetup{capbesideposition={right,top},capbesidewidth=4cm}}]{figure}[\FBwidth]
{\includegraphics[width=0.59\textwidth]{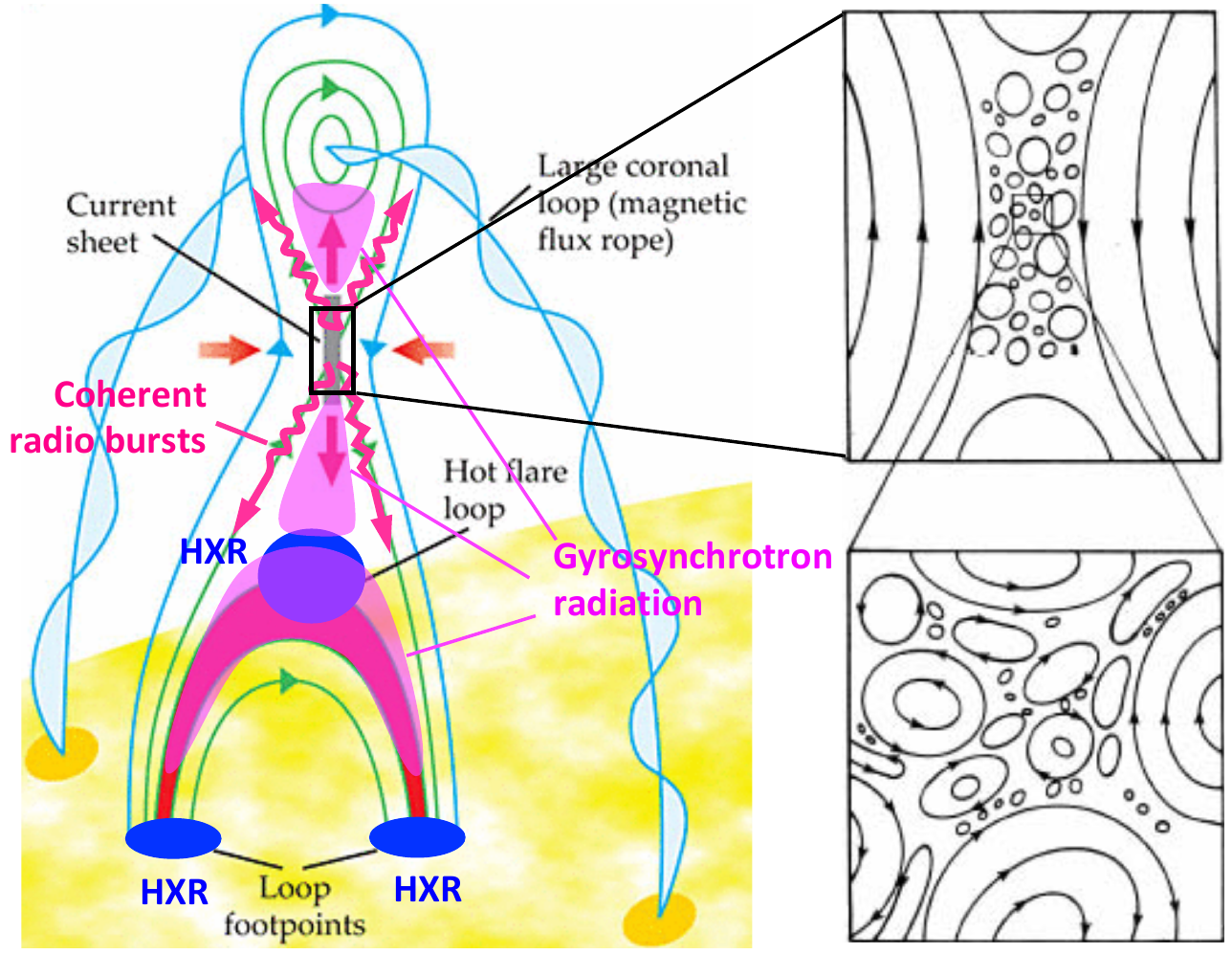}}
{\caption{Left: Schematic illustration of the standard model of eruptive solar flares (adapted from \cite{1995ApJ...451L..83S,2017arXiv170100792C}). Coherent radio bursts and incoherent gyrosynchrotron radiation can provide excellent diagnostics of the fragmentary magnetic reconnection and energy release processes (right panels adapted from \cite{2011LRSP....8....6S}.)}\label{fig:cartoon}}
\end{figure}

In order to release up to 10$^{32}$ ergs of energy in large X-class solar flares, a large coronal volume, sometimes comparable to a sizable fraction of a solar radius (10$^9$--10$^{10}$ cm), is needed to participate in the energy release process \cite{2011LRSP....8....6S}. The size of the flare energy release region can be estimated as
\begin{equation}
    L=\left(\frac{\varepsilon_{\rm flare}}{B^2/8\pi}\right)^{1/3}\approx 0.1R_{\odot} \left(\frac{\varepsilon_{\rm flare}}{10^{32}\ {\rm erg}}\right)^{1/3} \left(\frac{B}{100 \ \rm{G}}\right)^{-2/3},
\end{equation}
where $B$ is the magnetic field strength in the solar corona. Such a large amount of magnetic energy has to be released within a short period of 10$^3$--10$^4$ s (tens of minutes), which requires the energy release to proceed at a rate many orders of magnitude faster than that associated with the spontaneous diffusion of the coronal magnetic field in the flaring coronal volume. The size of the magnetic diffusion region for fast reconnection to proceed, in turn, has to be as small as $10^3$ cm. To bridge such a huge gap in spatial scales, flare models usually invoke a large-scale, but fragmented reconnection current sheet (CS; see Fig. \ref{fig:cartoon} for a schematic) \cite{2011LRSP....8....6S,2001EP&S...53..473S, 2006Natur.443..553D,2009PhPl...16k2102B}, or a large number of small-scale magnetic reconnection sites distributed throughout the flaring volume \cite{1994ISAA....1.....P}.

For decades we have observed solar flares with ever-increasing spatial, temporal, and/or spectral resolution at different wavelengths. Several models now exist to account for a variety of phenomenological aspects of flares under different reconnection geometries. One of the most well-known models is the CSHKP model (after \cite{1964NASSP..50..451C,1966Natur.211..695S,1974SoPh...34..323H,1976SoPh...50...85K}; also referred to as the \textit{standard model} of solar flares), which involves reconnection within a large-scale CS induced by an erupting magnetic flux rope (Fig. \ref{fig:cartoon}). 
However, until very recently, our knowledge of magnetic reconnection, the ``central engine'' of flares, was largely limited to what we could infer indirectly from observations of the morphology and dynamics of the newly reconnected magnetic flux tubes that are populated with thermal plasma bright in extreme ultraviolet (EUV) or soft X-ray (SXR) wavelengths. Now, at radio wavelengths, thanks to recent advances in radio interferometry and ultra-fast digital sampling and processing, radio observations have evolved from either Fourier synthesis imaging at a limited number of frequency channels or total-power dynamic spectroscopy to the ability to do both simultaneously, \textit{radio dynamic imaging spectroscopy}, which offers radio imaging with simultaneously high spectral and temporal resolution over a large number of contiguous frequency channels. Radio observations of this kind have begun to provide unique insights of the coronal magnetic field in the flaring volume and nonthermal electrons accelerated at or in close proximity to the reconnection region \cite{2013ApJ...763L..21C,2015Sci...350.1238C,2017ApJ...851..151M,2017A&A...606A.141R,2018ApJ...866...62C,2018ApJ...863...83G,2018A&A...611A..57M,2018NatSR...8.1676C}.

This white paper emphasizes exciting opportunities enabled by the new radio observing technique for studying \textit{magnetic reconnection and the associated magnetic energy release processes} in solar flares. We will first briefly review some outstanding challenges of this topic, and the observations required to address them. We then use two recent examples to demonstrate the science potential. Finally, we will discuss breakthroughs to this field that will be brought by a next-generation solar radio telescope such as the \textit{Frequency Agile Solar Radiotelescope}. Results from these studies, particularly when complemented by other multi-wavelength observations in optical, (E)UV, soft- and hard X-ray, and $\gamma$-ray wavelengths as well as numerical modeling, will provide unique insights into magnetic reconnection in many laboratory, space, and astrophysical plasma contexts.

\vspace{0.2cm}

\noindent \textbf{\large Outstanding Challenges}

A number of outstanding challenges exist in understanding the fundamental physical processes underlying magnetic reconnection and the associated magnetic energy release and conversion processes in various space, astrophysical, and laboratory plasma contexts. A summary of these challenges is provided in the white paper by \textbf{Ji \textit{et al.}} Here we outline three fundamental questions on which significant insights will be drawn from remote-sensing observations of solar flares:
\begin{enumerate}[wide, labelwidth=!, labelindent=0pt,itemsep=0.2pt,topsep=0pt]

\item  \textbf{\fontsize{13}{16}\selectfont Where and how does magnetic reconnection and energy release occur?} 

More specifically, what is the global magnetic and plasma context in which the reconnection occurs? 
To date, the location of magnetic reconnection has often been inferred indirectly from the geometry of newly reconnected magnetic flux tubes populated with thermal plasma \cite{2013NatPh...9..489S,2013Natur.493..501C,2014ApJ...797L..14T,2015ApJ...807....7R,2015NatCo...6E7598S,2018ApJ...854..122W}, or from coronal magnetic field models based on magnetograms measured at the photospheric/chromospheric level \cite{2013ApJ...778L..36L,2013ApJ...779..129K,2014ApJ...788..182I,2018ApJ...869...13J}. Hard X-ray (HXR) observations in the past two decades have provided critical insights on this matter, positioning the reconnection site somewhere above the top of reconnected flare arcades, at least for large eruptive flares \cite{1994Natur.371..495M,2008A&ARv..16..155K}. Yet the precise location of reconnection and its spatiotemporal evolution is largely unknown, due to, in part, the difficulty in tracing accelerated electrons from the reconnection site(s) with sufficient sensitivity and dynamic range. More direct and sensitive means for \textit{pinpointing the reconnection site(s)} is thus urgently needed. 

\item  \textbf{\fontsize{13}{16}\selectfont What is the physical nature of the reconnection site(s)?} 

The prevalence of fast subsecond-to-second-scale bursts in radio and HXR light curves of flares implies that the magnetic reconnection is probably highly fragmentary in both time and space \cite{2002SSRv..101....1A}. Yet it remains poorly known whether such fragmentation takes place locally, in a large-scale structure (such as the elongated CS in the standard flare model), or is widely distributed throughout the flare volume. Distinguishing these scenarios requires the \textit{capability of tracing the reconnection signatures with simultaneous high sensitivity and spatial resolution at subsecond time cadence}, which has not been routinely available. More importantly, direct identification of the nature of the reconnection sites requires the knowledge of its \textit{inherent magnetic properties}, which has been hitherto missing due to the lack of measurements of the magnetic field in the solar corona. 

\item  \textbf{\fontsize{13}{16}\selectfont Where and how does the energy conversion occur?}

Once the fast magnetic reconnection is triggered, the inflowing magnetic energy in the form of Poynting flux is quickly converted into other forms of energy: accelerated particles, heated plasma, turbulence and waves, and bulk flows. The sites for particle acceleration and plasma heating, for example, do not necessarily coincide with the reconnection region, while their locations and the mechanisms responsible remain largely unknown \cite{1997JGR...10214631M,2011SSRv..159..357Z}. The detailed energy partition in different energy forms following magnetic reconnection is also poorly constrained by observational data, although efforts have been made on constraining the global (spatially and temporally integrated) energetics of large flares \cite{2012ApJ...759...71E,2014ApJ...797...50A,2016ApJ...832...27A}. Making progress on this matter requires the capability of quantifying the magnetic energy release in time and space, as well as the means for deriving the energy content of the accelerated particles, thermal plasma, and turbulence/waves.

\end{enumerate}

\vspace{0.2cm}

\noindent \textbf{\large New Opportunities at Radio Wavelengths}

Radio observations of flares are in a strong position to make revolutionary breakthroughs on each of the challenges above, taking advantage of two types of radio emission in solar flares (see left panel of Fig. \ref{fig:cartoon} for an illustration):
\begin{itemize}[noitemsep,topsep=0pt]
\item\textbf{Coherent radio bursts} serve as a highly sensitive means for tracing nonthermal electrons at or around the reconnection site, thereby probing important characteristics of reconnection including the location, its fragmentary nature, and the presence of reconnection-driven shocks, waves, and turbulence (see, e.g., \cite{2015Sci...350.1238C,2018ApJ...866...62C,2019ApJ...872...71Y} for recent examples).

\item\textbf{Incoherent gyrosynchrotron radiation} can be used to measure the evolving magnetic field at or around the reconnection site, and to quantify the distribution function of flare-accelerated nonthermal electrons (see, e.g., \cite{2018ApJ...863...83G,2013SoPh..288..549G,2014ApJ...787..125N,2017ApJ...843....8C} for recent examples).
\end{itemize}
\noindent Such diagnostics with both radio emission types call for routine, solar-dedicated radio observations of flares in a broad range from centimeter to meter wavelengths with sufficient spatial, spectral, and temporal resolution (i.e., \textit{``broadband radio dynamic imaging spectroscopy''}), combined with high dynamic range and polarization measurements. Currently there is no solar-dedicated radio instrument with these capabilities. However, the recent commissioning of two radio facilities, the general-purpose \textit{Karl G. Jansky Very Large Array (VLA)} and the 13-element solar-dedicated pathfinder array \textit{Expanded Owens Valley Solar Array (EOVSA)}, have already started to provide exciting demonstration science results. In the following we briefly discuss two examples to demonstrate the unique power of this radio imaging spectroscopy. 

\noindent  \textbf{\fontsize{13}{16}\selectfont Example 1: Pinpoint the fragmentary magnetic reconnection sites} 

Fast electron beams ($\sim$0.1--0.5c) accelerated at or in close proximity to the magnetic reconnection region emit a type of coherent radio emission---known as ``type III radio bursts'' (see \cite{2014RAA....14..773R} for a review)---when they propagate along newly reconnected magnetic field lines. 
Thanks to their coherent radiation nature, these bursts are usually much brighter than the background, making them an extremely sensitive means for tracing the electron beams. In order to trace the beams over a wide range of coronal heights to their site(s) of origin, a wide frequency bandwidth combined with dense spectral sampling are required to observe the type III bursts from decimetric to metric wavelengths. Furthermore, as the magnetic energy release proceeds in a highly fragmentary fashion, probably in both space and time \cite{2001EP&S...53..473S,2002SSRv..101....1A,2016ASSL..427..373S}, subsecond temporal cadence and subarcsecond spatial resolution of the emission centroids are crucial to distinguish different reconnection events. 

\begin{figure*}[!ht]
\begin{center}
\includegraphics[width=1.0\textwidth]{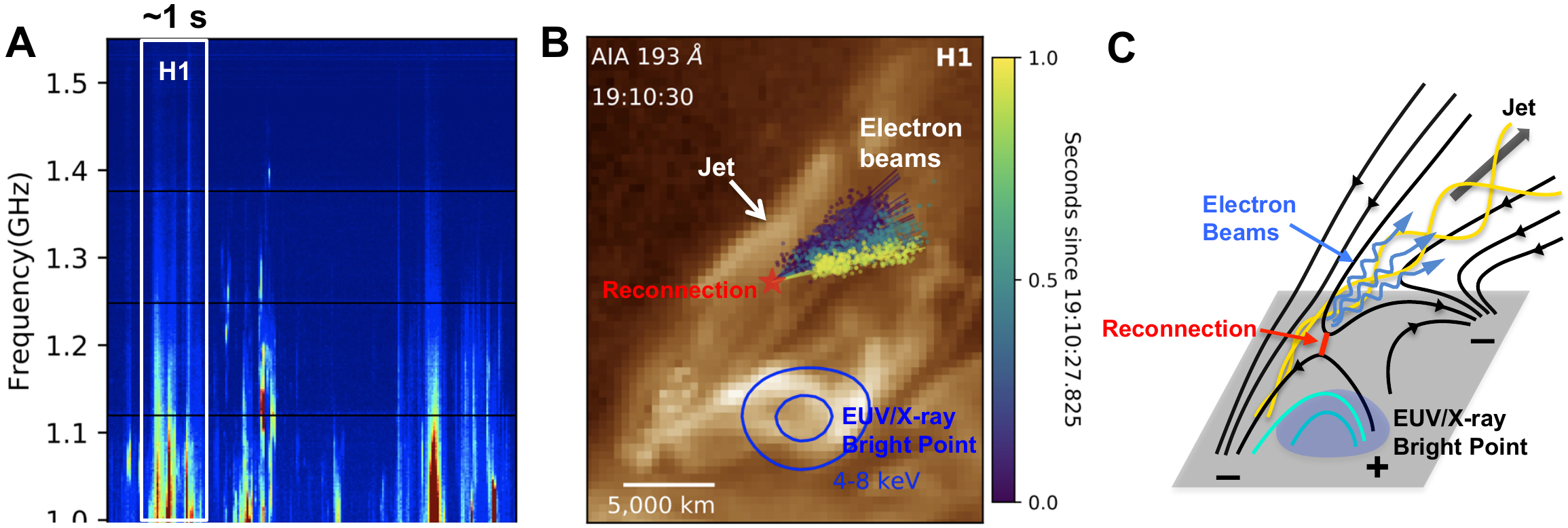}
\caption{Trajectories of semi-relativistic electron beams associated with a type III radio burst group of $\sim$1-s duration in a solar jet. (A) Type III bursts visible as the nearly vertical bright features in the radio dynamic spectrum. (B) Each trajectory (colored in time) derived from a burst in panel (A) within 0.05 s is delineated by a series of nearly linearly distributed source centroids. All the beams appear to emanate from an extremely compact, $<$600 km$^2$ region in the low solar corona, which is likely a macroscopic magnetic reconnection null point (adapted from \cite{2018ApJ...866...62C}.)}\label{fig:tp3}
\end{center}
\end{figure*}

Figure \ref{fig:tp3} shows a recent example from \textit{Jansky VLA} observations of type III radio bursts associated with a reconnection event that occurred with a solar jet. The observations were made in the 1--2 GHz frequency band with 50-ms temporal cadence and a centroid accuracy of $\sim$1'' ($\sim$700 km on the Sun). Multitudes of semi-relativistic electron beams ($>$0.5$c$) associated with a brief, $\sim$1-s-duration burst group are seen to diverge from an extremely compact region in the low corona, which corresponds to a macroscopic magnetic reconnection null point. The null point is likely highly fragmentary as each electron-beam-conducting magnetic field line displays very different position angle and inherent density properties. Moreover, different burst groups originate from distinct null points, suggesting these macroscopic null points are spatially (and temporally) distributed. We note, however, that the \textit{Jansky VLA} and the proposed \textit{Next Generation VLA (ngVLA)} are both general purpose facilities with very limited observing time for solar science as well as a restricted instantaneous bandwidth. Since it is not currently possible to predict when solar flares occur, only isolated cases recorded by these facilities, largely by chance, could be studied in detail. A solar-dedicated instrument with the necessary capabilities will provide routine observations to \textit{pinpoint the magnetic reconnection sites and probe their fragmentary nature for flares of all sizes and with a variety of reconnection geometries}.  

\noindent  \textbf{\fontsize{13}{16}\selectfont Example 2: Decipher the magnetic nature of reconnection sites} 

Microwave (MW; centimeter wavelengths) emission during flares is usually dominated by incoherent gyrosynchrotron radiation from mildly relativistic electrons (energies of a few 100s of keV to MeV) gyrating in the coronal magnetic field \cite{1998ARA&A..36..131B}. The MW spectra can be used to constrain the energy content of flare-accelerated electrons, which, particularly when combined with imaging spectroscopic observations in HXRs \cite{2017arXiv170100792C,2011SSRv..159..225W}, is invaluable for constraining particle acceleration and transport processes in flares. (See the white paper by \textbf{Gary \textit{et al.}}) Meanwhile, the MW spectra contain unique information about the magnetic field strength and direction in the source region. Spatially- and temporally-resolved measurements of the MW spectrum will enable us to quantify the magnetic field structure and evolution in and around the magnetic reconnection site. 

The approach of using broadband, spatially-resolved MW spectra to recover the coronal magnetic field and the nonthermal electron distribution along each line of sight in the flaring region has been demonstrated through simulated MW observations of a model flare loop model \cite{2013SoPh..288..549G}. More recently, by using MW imaging spectroscopic data from the newly commissioned \textit{EOVSA}, this technique was successfully applied to derive both the magnetic field strength and nonthermal electron distribution at selected pixels within the looptop region of a large solar flare \cite{2018ApJ...863...83G}. The same method can be used to derive these physical properties for every pixel in the image, as long as the observed microwave data have adequate number of spectral measurements with good signal-to-noise ratio that sample different portions of the gyrosynchrotron spectrum. 
The derived spatial variation and temporal evolution of the magnetic field around the reconnection site will enable direct comparison with model predictions to identify the magnetic nature of the reconnection sites, and moreover, to quantify the magnetic energy release and its spatiotemporal variation. We emphasize that, although \textit{EOVSA} has already demonstrated the great science potential of this new technique as a pathfinder for the next-generation solar radio telescope, it is a small array with only 13 elements. Its angular resolution is limited to $\sim$60$''$/GHz and its image dynamic range is limited to $\sim$100:1 at best. To measure the time-evolving magnetic field and flare-accelerated electron distribution in a much broader area around the magnetic reconnection site with sufficient spatial details, \textit{an instrument with  high-fidelity, high-dynamic-range, and high-resolution dynamic imaging spectroscopy is required}.

\vspace{0.2cm}

\noindent \textbf{\large Concluding Remarks}

The examples above are merely a very small subset of all possible diagnostics of magnetic reconnection and the associated energy release and conversion processes in solar flares enabled by the new technique of radio dynamic imaging spectroscopy. Breakthroughs on this topic call for a telescope with sufficient numbers of antennas to ensure high-fidelity, high-dynamic-range, and high-resolution imaging, combined with good spectral and temporal resolution over a wavelength range sufficient to sample both coherent radio bursts and incoherent gyrosynchrotron emission. The utility and impact of an instrument with this capability will be profound not only for the topic focused here, but also many other topics including particle acceleration, chromospheric and coronal magnetic field measurements, shocks and coronal mass ejections in the coronal and interplanetary space. We direct readers to white papers by \textbf{Bastian \textit{et al.}}, \textbf{Fleishman \textit{et al.}}, and \textbf{Gary \textit{et al.}} for discussions on these topics. An instrument with the attributes described above has already been defined: the \textit{Frequency Agile Solar Radiotelescope}. The concept is mature and it is ready for implementation. Separate project-oriented white papers will describe the instrument in detail.
\pagebreak
\bibliography{reconnection}

\bibliographystyle{ieeetr}

\end{document}